\documentclass[fp,twocolumn]{jpsj3}
\usepackage{amsmath}
\usepackage{color}

\title{Helical-to-Fan Transitions under Magnetic Fields in the Noncentrosymmetric Tetragonal Magnet EuRhGe$_3$}

\author{Takeshi Matsumura$^1$, Hiroto Nitta$^1$, Hironori Nakao$^2$, Masashi Kakihana$^3$, Masato Hedo$^3$, Takao Nakama$^3$, and Yoshichika \={O}nuki$^{3,4}$}
\inst{
$^1$Department of Quantum Matter, ADSE, Hiroshima University, Higashi-Hiroshima 739-8530, Japan \\
$^2$Photon Factory, Institute of Materials Structure Science, High Energy Accelerator Research Organization, Tsukuba, 305-0801, Japan \\
$^3$Faculty of Science, University of the Ryukyus, Nishihara, Okinawa 903-0213, Japan \\
$^4$RIKEN Center for Emergent Matter Science, Wako, Saitama 351-0198, Japan
} 

\abst{
The magnetic structure of EuRhGe$_3$, a noncentrosymmetric body-centered tetragonal magnet with the space group $I4mm$, has been investigated by resonant X-ray diffraction. Below $T_{\text{N}}=12$ K, EuRhGe$_3$ undergoes a helical magnetic ordering with an incommensurate propagation vector $\mib{q}=(0, 0, 0.809)$, in which the magnetic moments lie in the $ab$ plane and rotate by a constant turn angle of $145.8^{\circ}$ between adjacent layers. When a magnetic field is applied along the $a$ axis at 2 K, a second-harmonic $2q$ peak develops, indicating that the circular helix is gradually distorted into a helimagnetic soliton-lattice state, which eventually undergoes a lock-in transition to the commensurate structure with $q=0.8$ at 3.8 T. 
Above the subsequent phase boundary at 5 T, the helicity is lost, and a spin-flop $xyz$-fan (elliptic conical) state is realized, in which the moments oscillate predominantly along the $b$ axis but are accompanied by a small $c$-axis component. 
At higher fields, the system enters a conventional planar $xy$-fan phase without a $c$-axis component. 
EuRhGe$_3$ provides a prototypical example of a helimagnet that exhibits a full sequence of field-induced structures, evolving from a circular helix to a spin-flop $xyz$-fan (elliptic conical), and finally to a planar $xy$-fan structure, which has been theoretically predicted. 
(\today)
}

\begin{document}
\maketitle

\section{Introduction}
Emergence of nontrivial magnetic structures, such as skyrmion lattices and chiral soliton lattices composed of noncollinear or noncoplanar spiral structures, has stimulated extensive studies~\cite{Togawa23,Togawa12,Honda20,Muhlbauer09,Yu10,Yamasaki15,Seki12,Kezsmarki15}. 
An important aspect in understanding these states is the relationship between crystallographic symmetry and the microscopic mechanism responsible for their stabilization. 
In long period structures of $d$-electron systems, where the spin texture can be approximated as a continuous medium, the Dzyaloshinskii-Moriya (DM) type interaction arising from broken symmetry plays an essential role in generating the twisting force~\cite{Bogdanov89,Bogdanov94}. 
In contrast, in nanoscale short period structures of $f$-electron systems, incommensurate spiral often originate from competing exchange interactions of Ruderman-Kittel-Kasuya-Yosida (RKKY) type~\cite{Hayami17,Hayami21,Hayami22,Yambe22,Hayami24,Hayami26}. 

In $f$-electron systems, magnetic skyrmion lattices have been well established in the chiral magnet EuPtSi~\cite{Kaneko19,Tabata19,Matsumura24a}, the noncentrosymmetric polar magnet EuNiGe$_3$~\cite{Matsumura24b,Singh23}, and the centrosymmetric magnets EuAl$_4$, GdRu$_2$Si$_2$, Gd$_2$PdSi$_3$, and Gd$_3$Ru$_4$Al$_{12}$~\cite{Kurumaji19,Hirschberger19,Khanh20,Khanh22,Takagi22,Wood23}. 
A common feature of these compounds is the $4f^7$ ($S=7/2$, $L=0$) configuration, which is free from crystal-field anisotropy and allows emergence of a wide variety of self-organized magnetic structures. 
It should be noted, however, that pronounced magnetic anisotropies often develop in the ordered phases, reflecting the underlying magnetic structures. 
Another important aspect is the position of the propagation vector ($\mib{q}$)  in reciprocal space. 
In RKKY systems, it is determined by the wave vector at which the generalized susceptibility $\chi(\mib{q})$ takes its maximum. 
This is closely related with the Fermi-surface structure~\cite{Bouaziz22,Sarkar25,Arai26}. 
Recently, the microscopic origin of such finite-$\mib{q}$ magnetic ordering has also been discussed in terms of inter-orbital frustration in multi-orbital systems~\cite{Nomoto20}. 

EuTGe$_3$ (T=transition metal) compounds, which share the BaNiSn$_3$-type body-centered tetragonal structure (space group $I4mm$), exhibit a variety of magnetic properties based on the $4f^7$ ($S=7/2$, $L=0$) configuration of Eu$^{2+}$~\cite{Bednarchuk15,Maurya16,Kakihana17,Utsumi18}. 
Among them, the magnetic structures of EuIrGe$_3$ and EuNiGe$_3$ have been studied in detail~\cite{Matsumura22,Kurauchi23,Matsumura24b}. 
In EuIrGe$_3$, a longitudinal sinusoidally modulated order with $\mib{q}\sim (0, 0, 0.792)$ appears below $T_{\text{N}}=12$ K, followed by successive transitions to cycloidal phases at $T_{\text{N}}^{\;\prime}=7.0$ K and $T_{\text{N}}^{\;*}=5.0$ K. In the intermediate and low-temperature phases, the magnetic moments rotate within the [100]-[001] plane and the [110]-[001] plane, respectively, with propagation vectors $\mib{q}\sim (\delta, 0, 0.794)$ ($\delta \sim 0.017$) and $\mib{q}\sim (\delta', \delta', 0.798)$ ($\delta' \sim 0.012$), respectively. 
Interestingly, the $\mib{q}$-vector slightly tilts away from the [001] direction, thereby preserving the symmetry compatible with the  $I4mm$ space group. 

In EuNiGe$_3$, a helimagnetic order is realized at zero field with $\mib{q}=(\delta_1, \delta_2, 0)$ ($\delta_1 = 0.26$ and $\delta_2 = 0.053$ at 2 K). When a magnetic field is applied along the $c$ axis, a distorted triangular skyrmion-lattice phase appears, which can be described as a superposition of three helical components. 
Remarkably, the helicity of the zero-field helix, determined by a DM-type interaction, is reversed so that all three components acquire the same helicity in the triple-$\mib{q}$ structure, a phenomenon referred to as helicity unification. 
These results demonstrate a rich variety of magnetic structures realized in the EuTGe$_3$ family and motivate further studies of other EuTGe$_3$ compounds to elucidate the relationship between the electronic structure and magnetic ordering. 

\begin{figure}
\begin{center}
\includegraphics[width=8cm]{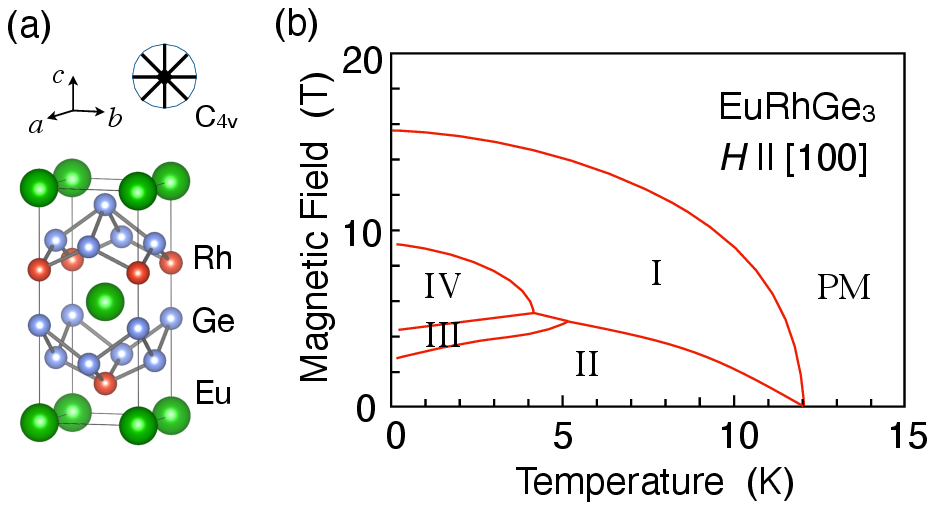}
\caption{(Color online) 
(a) The body-centered tetragonal structure of EuRhGe$_3$ (point group $C_{4v}$), which possesses a fourfold rotation axis along the $c$ axis and mirror planes parallel to the $c$ axis.  The crystal structure was drawn using the program VESTA~\cite{momma11}. 
(b) Magnetic phase diagram of EuRhGe$_3$ for $H \parallel [100]$ constructed from the bulk property measurements~\cite{Maurya16}. 
}
\label{fig:crysltSTMagPD}
\end{center}
\end{figure}

The physical properties of EuRhGe$_3$ have been well studied~\cite{Bednarchuk15,Maurya16,Kakihana17,Utsumi18,Utsumi21,Dhami24}. 
Antiferromagnetic order sets in below $T_{\text{N}}=12$ K. 
The magnetic susceptibility in the ordered phase indicates that the ordered moments lie in the $ab$ plane. 
When a magnetic field is applied along the [100] direction at low temperatures, several field-induced phase transitions occur before the fully polarized state is reached~\cite{Maurya16}. The crystal structure and the magnetic phase diagram for $H\parallel [100]$ is shown in Fig.~\ref{fig:crysltSTMagPD}, where the phase nomenclature follows Ref.~\citen{Maurya16}. 
The fact that the I--II and PM--I phase boundaries merge at zero field is reminiscent of the helical-to-fan transitions commonly observed in EuCo$_2$Ge$_2$, EuRh$_2$Ge$_2$, and EuIr$_2$Ge$_2$~\cite{Takeuchi21}. 
In magnetic fields, the paramagnetic (PM) phase is separated from the low-temperature ordered phases by phase I.
Below approximately 4 K, two additional intermediate phases, III and IV, appear between phases II and I.

The purpose of our study is to clarify the magnetic structures of the magnetic phases of EuRhGe$_3$ using resonant X-ray diffraction. 
The experimental procedures are described in \S \ref{sec:exp} and the Appendix. 
In \S \ref{sec:Results}, we show that the zero-field state is an incommensurate helimagnetic order with $\mib{q}=(0, 0, 0.809)$, in which the two in-plane magnetic components have equal amplitudes. 
When a magnetic field is applied along the $a$ axis at 2 K, a second harmonic $2q$ reflection develops, indicating that the circular helix is distorted into a helimagnetic soliton-lattice state, which eventually undergoes a lock-in transition to phase III with $\mib{q}=(0, 0, 0.8)$. 
Since both the magnetic helicity and the oscillation of the $a$-axis component disappear in phases I and IV, we investigate the difference between these two phases in \S \ref{sec:3-5} through an analysis of temperature dependence at 6 T.  
We conclude that phase IV is most likely to be a spin-flop $xyz$-fan (elliptic conical) state, in which the moments oscillate in the $ab$ plane and are accompanied by a small $c$-axis component. Phase I, in contrast, is most likely a conventional planar $xy$-fan state without a $c$-axis component. 
Finally in \S \ref{sec:discuss}, we discuss the evolution of the magnetic structures and show that EuRhGe$_3$ provides a prototypical example of a helimagnet exhibiting a full sequence of field-induced phases, evolving from a circular helix to a spin-flop $xyz$-fan (elliptic conical), and finally to a planar $xy$-fan state, which has been theoretically predicted~\cite{Nagamiya62,Johnston19,Johnston17}.

\section{Experiment}
\label{sec:exp}
A single crystal of EuRhGe$_3$ was grown by the In-flux method~\cite{Kakihana17}. 
Resonant X-ray diffraction (RXD) experiments were performed at BL-3A of the Photon Factory, KEK, Japan. 
The scattering geometry is shown in the Appendix. Measurements were performed at  X-ray energies around the Eu $L_2$ absorption edge. 
A plate-shaped sample with a mirror-polished (001) surface, 2.0$\times$1.5 mm$^2$ in area and 0.43 mm in thickness, was mounted in a vertical field 8 T superconducting cryomagnet so that the magnetic field was applied along the $a$ axis and the scattering vector lay in the $bc$-plane. 
We have not distinguished between the $[001]$ and $[00\bar{1}]$ directions, although they are not crystallographically equivalent in this noncentrosymmetric crystal structure. 
No significant difference was observed between measurements performed on the two opposite (001) surfaces.  

To investigate the magnetic helicity of the spiral spin arrangement, a phase retarder system was employed, which enabled us to control the incident X-ray polarization to right-handed circular polarization (RCP) or left-handed circular polarization (LCP). By analyzing the scattering intensity as a function of the incident polarization state, we investigated the magnetic helicity of the ordered structure~\cite{Matsumura24b,Kurauchi23}. 
To determine the magnetic Fourier component more precisely, we also performed a linear polarization analysis for the $\pi$-polarized incident beam using the 006 Bragg reflection of a pyrolytic graphite (PG) analyzer crystal. 
The analyzer $2\theta_{\text{A}}$ angle was 93.5$^{\circ}$ at the resonance energy of 7.613 keV, allowing efficient discrimination between the $\sigma'$ and $\pi'$ polarization components, which were measured at $\phi_{\text{A}}=0^{\circ}$ and $90^{\circ}$, respectively.

\section{Results and Analysis}
\label{sec:Results}
\subsection{Incommensurate magnetic order at zero field}
\label{sec:3-1}

\begin{figure}[t]
\begin{center}
\includegraphics[width=8.5cm]{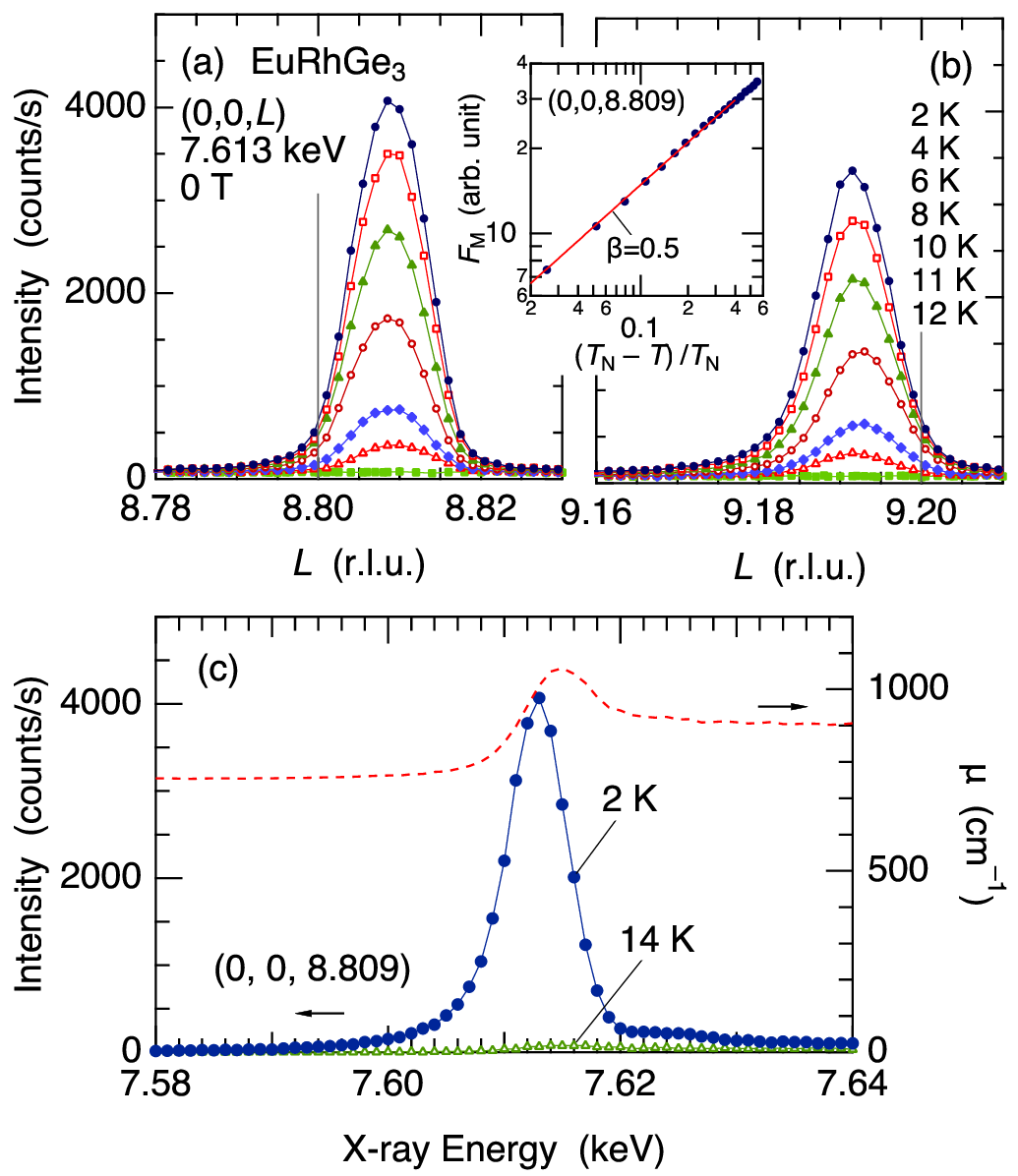}
\caption{(Color online) 
(a,b) Temperature dependence of the peak profiles along $(0, 0, L)$ through the $(0, 0, 8+q)$ and $(0, 0, 10-q)$ magnetic reflections at the resonance energy of 7.613 keV in zero field. The inset shows the temperature dependence of the magnetic structure factor $F_{\text{M}}$ (square root of the intensity) in arbitrary units as a function of the reduced temperature $(T_{\text{N}}-T)/T_{\text{N}}$. The solid line represents a fit using a mean-field critical exponent $\beta=0.5$. 
(c) X-ray energy dependence of the intensity of the $(0, 0, 8.809)$ magnetic reflection at 2 K in the ordered phase and at 14 K in the paramagnetic phase. The dashed line represents the absorption coefficient derived from the fluorescence spectrum. 
}
\label{fig:TEdep0T}
\end{center}
\end{figure}

\begin{fullfigure}[t]
\begin{center}
\includegraphics[width=15cm]{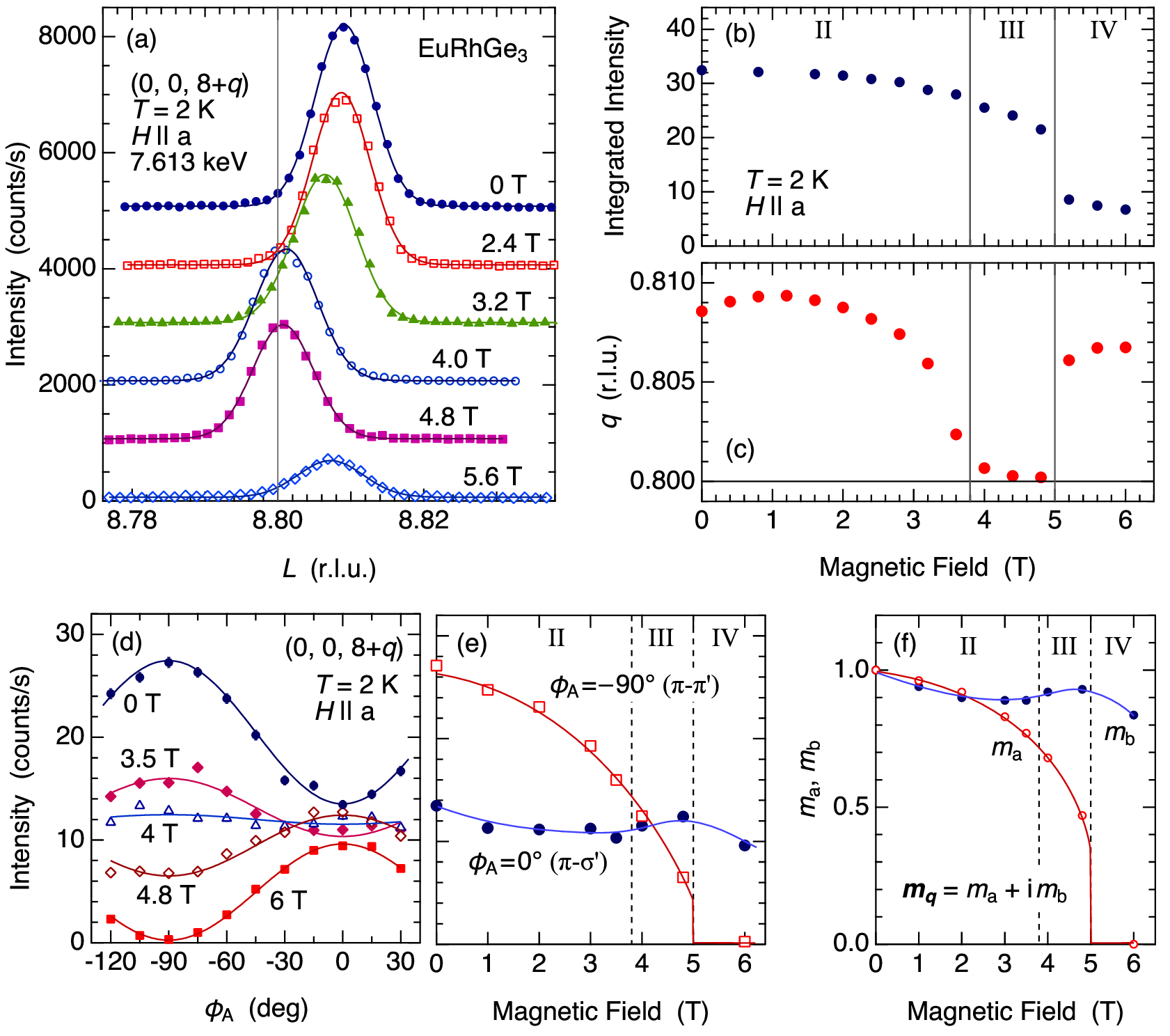}
\caption{(Color online) 
(a) Magnetic-field dependence of the $L$-scan profiles for $(0, 0, 8+q)$ at the lowest temperature of 2 K without polarization analysis. The vertical line represents the commensurate position of $q=0.8$. 
(b) Magnetic-field dependence of the integrated intensity for the $L$ scan in (a). The vertical lines represent the phase boundaries. 
(c) Magnetic-field dependence of the $q$-value. 
(d) Analyzer angle ($\phi_{\text{A}}$) dependence of the resonant peak intensity in magnetic fields. $\phi_{\text{A}}=0^{\circ}$ and $-90^{\circ}$ correspond to the $\pi$-$\sigma'$ and $\pi$-$\pi'$ channels, respectively. The solid lines are the fits to the data. 
(e) Magnetic field dependence of the resonant peak intensity at $\phi_{\text{A}}=0^{\circ}$ ($\pi$-$\sigma'$) and $\phi_{\text{A}}=-90^{\circ}$ ($\pi$-$\pi'$). 
(f) Magnetic field dependence of the magnetic Fourier components along the $b$ axis ($m_b \perp H$) and the $a$ axis ($m_a \parallel H$) deduced from the data in (e). The solid lines in (e) and (f) are guides to the eye.  
}
\label{fig:MagFdep}
\end{center}
\end{fullfigure}

Figures \ref{fig:TEdep0T}(a) and \ref{fig:TEdep0T}(b) show $L$ scans through the $(0, 0, 8+q)$ and $(0, 0, 10-q)$ magnetic reflections, respectively, illustrating the evolution of the magnetic Bragg peaks below $T_{\text{N}}$ with decreasing temperature.  
These scans clearly demonstrate that the magnetic order at zero field in phase II is characterized by an incommensurate propagation vector of $q=0.809$, which deviates from the commensurate value of 0.8. 
The peak position remains nearly constant down to the lowest temperature. This contrasts with the behavior in EuNiGe$_3$ and EuIrGe$_3$, where the $\mib{q}$ vector varies with temperature~\cite{Matsumura24b,Matsumura22,Kurauchi23}. 
The inset shows the $T$ dependence of the square root of the intensity ($\propto F_{\text{M}}$) plotted as a function of the reduced temperature $(T_{\text{N}}-T)/T_{\text{N}}$. The data are well described by a mean-field behavior with a critical exponent $\beta=0.5$.  
Figure  \ref{fig:TEdep0T}(c) shows the X-ray energy dependence of the magnetic Bragg peak at $(0,0,8.809)$. A pronounced resonance is observed at 7.613 keV, corresponding to the Eu-$L_2$ absorption edge, confirming that the magnetic Bragg peak originates from the ordered Eu moments.

\subsection{Magnetic field $\parallel$ $a$-axis}
\label{sec:3-2}
The magnetic-field evolution for $H \parallel a$, including the results of linear polarization analysis, is summarized in Fig.~\ref{fig:MagFdep}. 
Figure \ref{fig:MagFdep}(a) shows the evolution of the $(0,0,8+q)$ magnetic Bragg peak with magnetic field at 2 K. 
With increasing magnetic field, the peak position gradually shifts toward the commensurate value $q=0.8$. 
Remarkably, in phase III between 4 and 5 T, the propagation vector exhibits lock-in behavior, strongly suggesting a transition to a commensurate state with $q=0.8$. 
In phase IV above 5 T, the peak position returns to the original incommensurate value. 
The field dependences of the integrated intensity and the peak position are summarized in Figs.~\ref{fig:MagFdep}(b) and \ref{fig:MagFdep}(c), respectively. 
The intensity gradually decreases up to 5 T and exhibits a sudden drop at the transition into phase IV, where the $q$-vector becomes  incommensurate again. In contrast, the intensity changes continuously across the II--III phase boundary. 

Figure \ref{fig:MagFdep}(d) shows the results of linear polarization analysis of the scattered X-rays. 
Here, $\phi_{\text{A}}$ denotes the analyzer angle measured from the horizontal scattering plane (see Appendix); 
The positions $\phi_{\text{A}}=0^{\circ}$ and $-90^{\circ}$ correspond to the $\pi$--$\sigma'$ and $\pi$--$\pi'$ channels, respectively. 
The measured intensity exhibits a maximum and a minimum at these positions. 
The scattering amplitude from the ordered magnetic moments can be expressed as 
\begin{equation}
F_{\varepsilon\varepsilon'}=(\mib{\varepsilon}^{\prime *}\times\mib{\varepsilon})\cdot \mib{m}_{\mib{q}} \;,
\end{equation}
where $\mib{\varepsilon}$ and $\mib{\varepsilon}'$ are the polarization vectors of the incident and scattered X-rays, respectively, and $\mib{m}_{\mib{q}}$ is the Fourier component of the magnetic structure. The data at zero field are well reproduced by assuming $\mib{m}_{\mib{q}}=\mib{\hat{x}} \pm i \mib{\hat{y}}$, corresponding to a helical magnetic structure with moments rotating in the $ab$ plane. The calculated $\phi_{\text{A}}$ dependence is shown by the solid line for 0 T in Fig.~ \ref{fig:MagFdep}(d).

This result, together with the helicity measurements described in Sec. \ref{sec:3-3}, strongly suggests that the zero-field magnetic structure is an $xy$ planar helix. 
This interpretation is supported by the fact that the Fourier component $\mib{m}_{\mib{q}}=\mib{\hat{x}} \pm i \mib{\hat{y}}$ belongs to a two-dimensional irreducible representation for $\mib{q}=(0, 0, \zeta)$ in the $I4mm$ space group~\cite{Matsumura22}. It is also consistent with the magnetic susceptibility measurements, which indicate that the ordered moments lie in the $ab$ plane~\cite{Maurya16,Kakihana17}.

The $\pi$-$\pi'$ intensity demonstrates the existence of the $a$-axis Fourier component. By symmetry, the $b$-axis component should also exist, giving rise to the observed $\pi$-$\sigma'$ intensity. 
Although a $c$-axis component would also contribute to the $\pi$-$\sigma'$ intensity and therefore cannot be excluded experimentally from the polarization analysis alone, the Fourier component $\mib{m}_{\mib{q}}=(0,0,1)$ belongs to a different irreducible representation and cannot coexist with the $xy$ helix at zero field within the $I4mm$ symmetry. The magnetic susceptibility below $T_{\text{N}}$ is also inconsistent with the presence of a $c$-axis component. 

We therefore adopt the $xy$ helix as the zero-field magnetic structure and use it as the starting point for the subsequent analysis in magnetic fields,  where the appearance of a $c$-axis component becomes symmetry-allowed. 
In phases II and III, the experimental results are consistently explained without introducing a $c$-axis component. 
The possible emergence of a small $c$-axis component becomes important only in distinguishing the phases I and IV, and is discussed separately in Sec. \ref{sec:3-5}. 

Figure \ref{fig:MagFdep}(e) shows the magnetic-field dependence of the $\pi$--$\pi'$ and $\pi$--$\sigma'$ intensities, which are directly related to the components of $\mib{m}_{\mib{q}}$ parallel ($m_a$) and perpendicular ($m_b$) to the magnetic field ($H \parallel a$), respectively. 
With increasing magnetic field, the $\pi$--$\pi'$ intensity decreases and becomes smaller than the $\pi$-$\sigma'$ intensity above 4 T. 
Both intensities vary continuously up to the phase III--IV boundary at 5 T, where the $\pi$--$\pi'$ intensity abruptly vanishes. This behavior corresponds to the discontinuous drop in the intensity observed at 5 T in Fig.~\ref{fig:MagFdep}(b). 
By expressing the magnetic Fourier component as $\mib{m}_{\mib{q}}=m_a \mib{\hat{x}} \pm i m_b \mib{\hat{y}}$, the parameters $m_a$ and $m_b$ can be extracted from fits to the data in Fig.~\ref{fig:MagFdep}(d). 
The resulting field dependences are shown in Fig.~\ref{fig:MagFdep}(f), starting from the circular helix at zero field. 
The vanishing of $m_a$ indicates that only the component perpendicular to the applied field survives in phase IV above 5 T, suggesting a transition to a fan structure.

\begin{figure}
\begin{center}
\includegraphics[width=8.5cm]{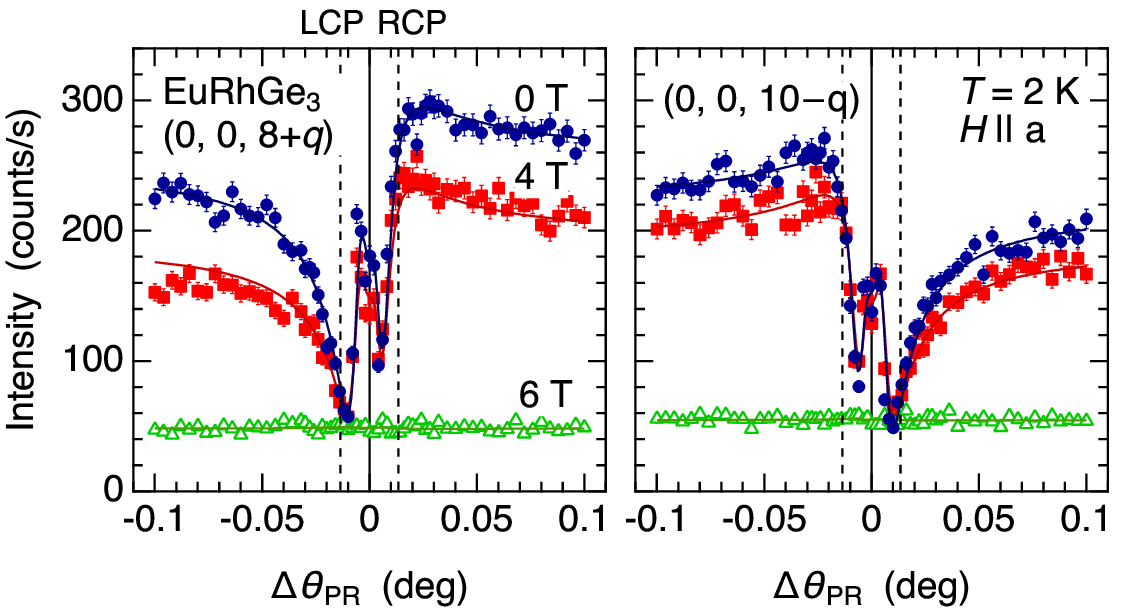}
\caption{(Color online) 
$\Delta\theta_{\text{PR}}$ dependence of the resonant peak intensities for $(0, 0, 8+q)$ and $(0, 0, 10-q)$ measured at 0 T, 4 T and 6 T at 2 K.  The background has been subtracted. The dashed lines indicate the LCP and RCP positions. 
The solid lines represent calculated intensity curves using the Fourier components shown in Fig.~\ref{fig:MagFdep}(f) and a mixture of $+$ and $-$ helicity domains with a ratio of 9:1. 
}
\label{fig:PRdep}
\end{center}
\end{figure}

\subsection{Magnetic helicity}
\label{sec:3-3}
Figure \ref{fig:PRdep} shows the incident-polarization dependence of the $(0, 0, 8+q)$ and $(0, 0, 10-q)$ magnetic reflections, where 
$\Delta\theta_{\text{PR}}=\theta_{\text{PR}} - \theta_{\text{B}}$ represents the offset angle of the diamond phase plate from the Bragg angle $\theta_{\text{B}}$ of the 111 reflection. 
The intensities clearly exhibit an asymmetric dependence between the LCP and RCP sides, indicating that the ordered structure possesses a magnetic helicity. 
The solid lines represent calculations based on the magnetic Fourier components $m_a$ and $m_b$ shown in Fig.~\ref{fig:MagFdep}(f). 
It should be noted, however, that the calculated curves assuming a single helicity, $\mib{m}_{\mib{q}}=m_a \mib{\hat{x}} + i m_b \mib{\hat{y}}$, show slight deviations from the experimental data. To reproduce the data more accurately, it was necessary to superpose a contribution from the opposite helicity domain, $\mib{m}_{\mib{q}}=m_a \mib{\hat{x}} - i m_b \mib{\hat{y}}$. 
The solid lines in Fig.~\ref{fig:PRdep} were obtained by assuming a mixture of the $+$ and $-$ helicity with a ratio of $9:1$. 

This behavior differs from that observed in EuIrGe$_3$ and EuNiGe$_3$, where the helicity in each domain is uniquely determined~\cite{Matsumura24b,Kurauchi23}. 
The large imbalance observed in EuRhGe$_3$ appears to be accidental, because measurements performed on a different region of the sample yielded a much weaker asymmetry, corresponding to a helicity ratio of approximately $6:4$. 
On the one hand, these results provide clear evidence that EuRhGe$_3$ exhibits a helical magnetic structure with a well defined helicity in each domain. 
On the other hand, they also suggest that this helix probably does not possess a preferred helicity. 
This is naturally expected for the present helical order, where the propagation vector is parallel to the fourfold $c$ axis and lies within mirror planes. 
In this geometry, the DM-type antisymmetric interaction vanishes~\cite{Yambe22}, allowing the $+$ and $-$ helicities to occur with equal probability. 

The data at 4 T show that the original helicity at zero field is preserved in the lock-in phase III. 
The $a$-axis Fourier component $m_a$ remains finite, indicating that the spins continue to rotate within the $ab$ plane.  
At 6 T in phase IV, however, as expected from the disappearance of $m_a$, the helical nature is completely lost, resulting in an almost flat $\Delta\theta_{\text{PR}}$ dependence. 

\begin{figure}
\begin{center}
\includegraphics[width=8.5cm]{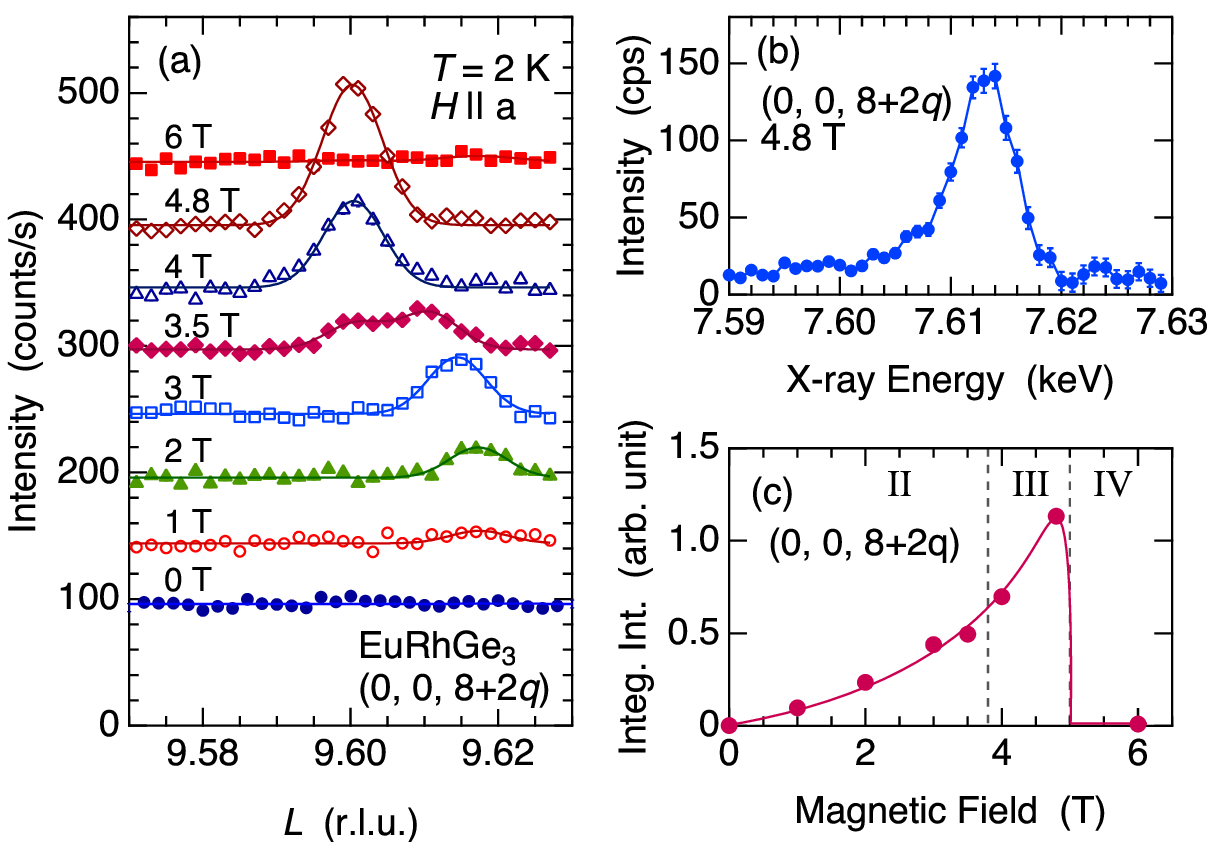}
\caption{(Color online) 
(a) Magnetic-field dependence of the peak profiles along $(0, 0, L)$ through the $(0, 0, 8+2q)$ reflection at 2 K. Solid lines are the fits to the data using Gaussian functions. 
(b) X-ray energy dependence of the $2q$ peak at 4.8 T. 
(c) Magnetic-field dependence of the integrated intensity of the $2q$ peak. The solid line is a guide to the eye. 
}
\label{fig:twoq}
\end{center}
\end{figure}

\subsection{Appearance of the $2q$ peak}
\label{sec:3-4}
When a magnetic field is applied perpendicular to the helical axis, the constant turn angle between magnetic moments on adjacent layers is modified. For example, in the chiral helimagnet Yb(Ni,Cu)$_3$Al$_9$, a periodic array of twisted spiral regions is formed, giving rise to a chiral soliton lattice (CSL)~\cite{Matsumura17}. 
Figure ~\ref{fig:twoq}(a) shows the magnetic-field dependence of the peak profile of the $(0,0,8+2q)$ reflection.  
No intensity is observed at zero field. 
With increasing magnetic field, the $2q$ intensity gradually develops, reaches its maximum at 4.8 T just before the transition to phase IV, and abruptly disappears above 5 T. 
At 3.5 T, immediately before the transition into the lock-in phase III, a coexistence of incommensurate and commensurate peaks is observed. 
Figure ~\ref{fig:twoq}(b) shows the X-ray energy dependence of the $2q$ peak at 4.8 T, which exhibits a clear resonance, indicating that the $2q$ peak is of magnetic origin rather than arising from a lattice distortion with half the period of the helimagnetic structure. 
Figure~\ref{fig:twoq}(c) shows the field dependence of the integrated intensity. The observed behavior closely resembles the field evolution of the $2q$ peak accompanying the formation of the CSL in Yb(Ni,Cu)$_3$Al$_9$. 

These results show that a soliton-like deformation of the helix is induced by a magnetic field applied perpendicular to the helical axis. 
In nonchiral helimagnets such as EuRhGe$_3$, the helix is expected to undergo a transition to a fan structure above a critical field. 
Below this critical field, however, such a nonlinear deformation of the helix is also allowed, even in the absence of crystallographic chirality.

\begin{figure}[t]
\begin{center}
\includegraphics[width=8.5cm]{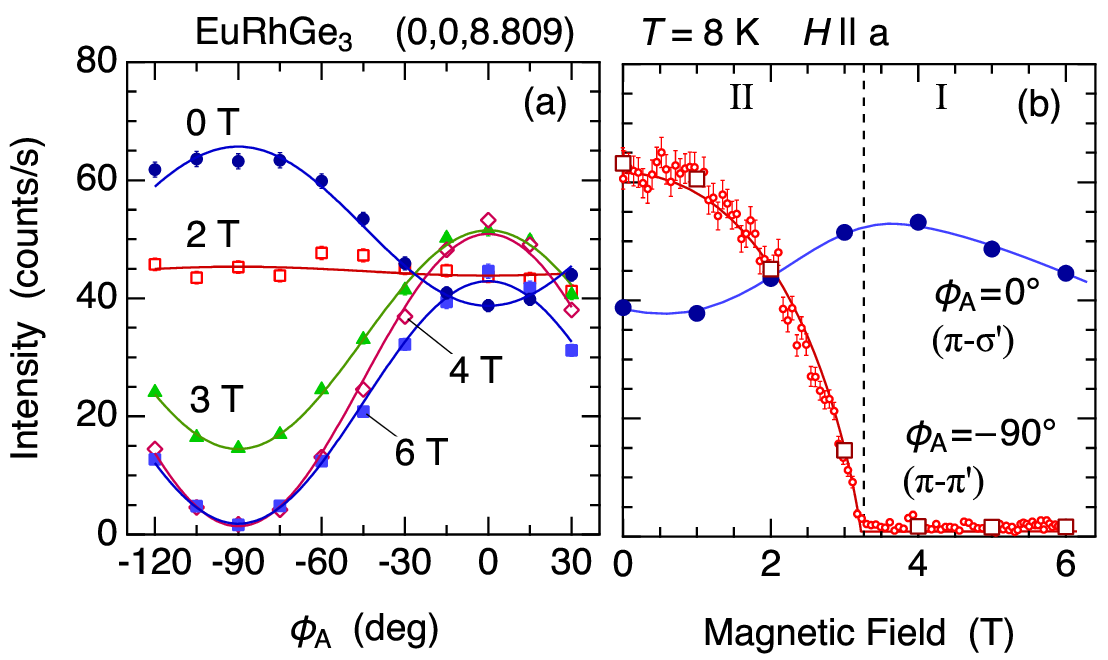}
\caption{(Color online) 
(a) Analyzer angle ($\phi_{\text{A}}$) dependence of the resonant peak intensity in magnetic fields at 8 K. 
(b) Magnetic-field dependence of the resonant peak intensity at $\phi_{\text{A}}=0^{\circ}$ ($\pi$-$\sigma'$) and $\phi_{\text{A}}=-90^{\circ}$ ($\pi$-$\pi'$) at 8 K. The solid lines are guides to the eye. The open circles represent the data obtained from a field-sweep measurement at $(0, 0, 8.809)$. 
}
\label{fig:pol8K}
\end{center}
\end{figure}

\subsection{Relation between phase I and phase IV}
\label{sec:3-5}
When a magnetic field is applied at 8 K, the ordered phase undergoes a direct transition from the helimagnetic phase II to the field-induced phase I, which appears at high fields above phases III and IV at low temperatures. 
Figure~\ref{fig:pol8K} shows the results of linear polarization analysis performed at 8 K in magnetic fields up to 6 T. 
The $\pi$--$\pi'$ intensity at $\phi_{\text{A}}=-90^{\circ}$ decreases smoothly to zero at the II--I phase boundary near 3 T, indicating the disappearance of $m_a$. Only the magnetic component perpendicular to the applied field remains in phase I. 
As shown in Fig.~\ref{fig:pol8K}(b), the magnitude of this perpendicular component increases slightly around the phase boundary. This is similar to the behavior observed at 2 K, where $m_a$ disappears on entering phase IV.

In both phases I and IV, the Fourier component parallel to the magnetic field, $m_a$, vanishes. 
Although the remaining component is most likely $m_b$, we do not have direct experimental evidence that $m_c$ is absent. 
At zero field, the polarization and helicity analyses at 2 K demonstrate a helimagnetic order with $m_a=m_b$. 
This equality is broken in magnetic fields, and $m_a$ vanishes in both phases IV and I. If only $m_b$ remains and $m_c=0$, the structure corresponds to a planar fan in which the magnetic moments oscillate only within the $ab$ plane. 
However, a well-defined phase boundary exists between the phases I and IV, as established by magnetization and resistivity measurements~\cite{Maurya16}. 
Therefore, the magnetic structures in these two phases must be different. 
Since only the $\pi$--$\sigma'$ intensity remains in both phases, it is difficult to distinguish between them from magnetic-field dependence measurements alone. 
To clarify the difference between phases I and IV, we measured the $T$ dependence of the $\pi$--$\sigma'$ intensity at 6 T. 

\begin{figure}[t]
\begin{center}
\includegraphics[width=7.5cm]{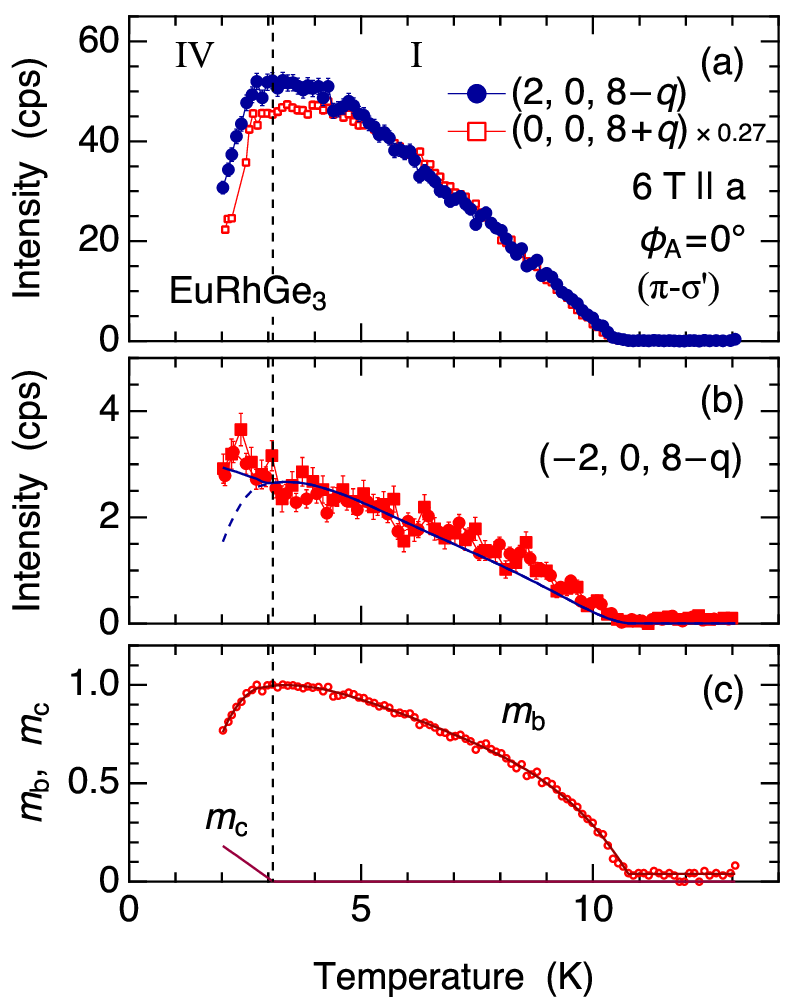}
\caption{(Color online) 
(a) Temperature dependence of the $(2, 0, 8-q)$ and $(0, 0, 8+q)$ intensities for the $\pi$--$\sigma'$ channel at 6 T. 
The data for $(0, 0, 8+q)$ are multiplied by 0.27. 
(b) Temperature dependence of the $(-2, 0, 8-q)$ intensity for $\pi$--$\sigma'$ at 6 T. 
(c) Temperature dependence of the Fourier component $m_b$ deduced from the $(2, 0, 8-q)$ data in (a). 
The dashed line in (b) shows the intensity expected from $m_b$ alone. The solid line in (b) is calculated by additionally assuming a finite $c$-axis component in phase IV, as shown by the solid line in (c). 
}
\label{fig:Tdep6T}
\end{center}
\end{figure}

Figure \ref{fig:Tdep6T} shows the $T$ dependences of the $\pi$--$\sigma'$ intensities of the $(0, 0, 8+q)$, $(2, 0, 8-q)$, and $(-2, 0, 8-q)$ reflections at 6 T. 
These reflections have different scattering geometries. Since $\mib{\varepsilon}_{\sigma}' \times \mib{\varepsilon}_{\pi}$ is parallel to the incident X-ray wave-vector $\mib{k}$, the scattering amplitude is proportional to the magnetic Fourier component parallel to $\mib{k}$. 
Accordingly, the $(0, 0, 8+q)$ intensity is sensitive to $m_b$ and $m_c$ with nearly equal weight, whereas 94\% of the $(2, 0, 8-q)$ intensity originates from $m_b$ and 96\% of the $(-2, 0, 8-q)$ intensity originates from $m_c$. 
Since the $(0, 0, 8+q)$ and $(2, 0, 8-q)$ intensities exhibit nearly identical $T$ dependences as shown in Fig.~\ref{fig:Tdep6T}(a), it can be concluded that both reflections are dominated by the $m_b$ component. 
The steep decrease in intensity below 3 K indicates that $m_b$ is reduced on entering phase IV. 
By taking the square root of the $(2, 0, 8-q)$ intensity, we extracted the $T$ dependence of $m_b$ as shown in Fig.~\ref{fig:Tdep6T}(c). 

The weak intensity observed at $(-2, 0, 8-q)$ in Fig.~\ref{fig:Tdep6T}(b) can be explained by taking into account the 4\% contribution from $m_b$. The dashed line in Fig.~\ref{fig:Tdep6T}(b) shows the $T$ dependence expected from the $m_b$ component alone, which reproduces the data in phase I. 
However, in phase IV, if only $m_b$ is present, the intensity is expected to decrease steeply as indicated by the dashed line. The experimental data in Fig.~\ref{fig:Tdep6T}(b), however, do not exhibit such a decrease; instead, the intensity remains nearly constant or even shows a slight increase across the boundary. 
One possible explanation is that an additional $m_c$ component develops in phase IV, with a magnitude as shown in Fig.~\ref{fig:Tdep6T}(c). The appearance of $m_c$ enhances the intensity of the $(-2, 0, 8-q)$ reflection and provides a good description of the observation.  
Although this analysis does not constitute a rigorous proof of the existence of $m_c$, it is consistent with the existence of a well-defined  boundary between phases I and IV.


\begin{fullfigure}[t]
\begin{center}
\includegraphics[width=17cm]{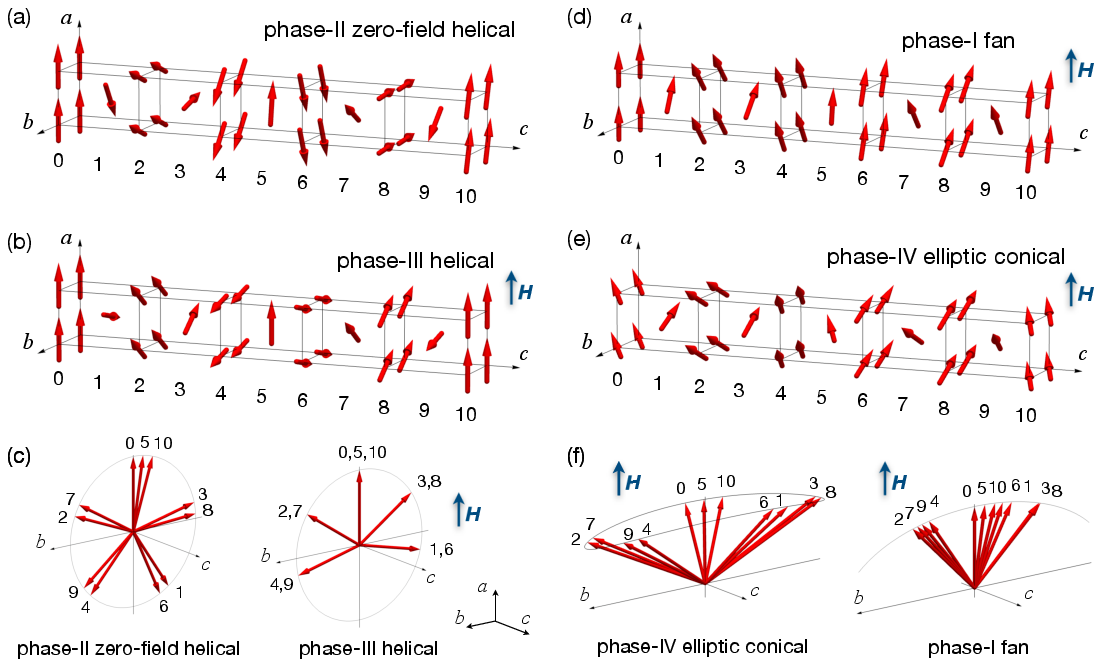}
\caption{(Color online) 
Illustrations of the magnetic structures of EuRhGe$_3$ at 2 K for the first five unit cells.  The numbers denote the layer index. 
(a) Helical magnetic structure at zero field with $\mib{q}=(0, 0, 0.81)$. The magnetic moments rotate by a constant turn angle of 145.6$^{\circ}$ between adjacent layers along the $c$ axis. 
(b) Lock-in state in phase III at 4.8 T with $\mib{q}=(0, 0, 0.8)$. A uniform ferromagnetic component corresponding to 33 \%  of the full moment (2.3 $\mu_{\text{B}}$/Eu) is assumed. The structure is drawn assuming an equal-moment configuration. 
(c) Distribution of magnetic-moment directions corresponding to the structures in (a) and (b).  
(d) Schematic illustration of the planar $xy$-fan structure in phase I, assuming 10 T, $\mib{q}=(0, 0, 0.81)$, and a uniform ferromagnetic component corresponding to 71 \%  of the full moment  (5 $\mu_{\text{B}}$/Eu). 
(e) Schematic illustration of the elliptic-conical ($xyz$-fan) structure in phase IV, assuming 6 T, $\mib{q}=(0, 0, 0.81)$, and a uniform ferromagnetic component corresponding to 43 \%  of the full moment  (3 $\mu_{\text{B}}$/Eu). 
(f) Distribution of the magnetic-moment directions corresponding to the structures in (d) and (e). 
The sparse regions are eventually filled uniformly when a sufficiently large number of unit cells is considered. 
}
\label{fig:Magst}
\end{center}
\end{fullfigure}

\section{Discussion}
\label{sec:discuss}
\subsection{Phases II and III}
The zero-field helical magnetic structure at the lowest temperature is shown in Fig.~\ref{fig:Magst}(a). The magnetic moments rotate within the $ab$ plane with equal amplitudes of $m_a$ and $m_b$, propagating along the $c$ axis with an incommensurate wave vector $\mib{q}=(0, 0, 0.809)$. The turn angle between the moments in adjacent Eu layers is 145.6$^{\circ}$. The moment directions in the first five unit cells are illustrated in Fig.~\ref{fig:Magst}(c). 
Since the $\mib{q}$ vector is slightly incommensurate with the lattice periodicity, the magnetic moments eventually sweep through all directions in the $ab$ plane, consistent with the weak magnetic anisotropy within the plane. 

The equal-amplitude helix, described by $\mib{q} \parallel c$ and two orthogonal Fourier components lying in the $ab$ plane, is compatible with an irreducible representation of the space group $I4mm$~\cite{Matsumura22}. This is also consistent with the absence of any detectable lattice distortion that would lower the tetragonal symmetry within the accuracy of the present X-ray diffraction experiment. 

The magnetic structure at 4.8 T in phase III, which is locked into a commensurate period corresponding to four turns in five unit cells, is shown in Fig.~\ref{fig:Magst}(b). Although the equal-amplitude helix is distorted in magnetic fields applied along the $a$ axis, Fig.~\ref{fig:Magst}(b) is drawn assuming an equal-moment structure. 
In this state, the magnetic moments still form a helical arrangement and rotate within the $ab$ plane, but the turn angle is no longer uniform. The moments spend more time aligned with the field direction and rapidly pass through the opposite orientation. Although this state may simply be described as a distorted helix, it may also be regarded as a helimagnetic soliton-lattice state by analogy with the CSL realized in chiral helimagnets. It should be noted, however, that EuRhGe$_3$ is not a chiral magnet. Moreover, the zero-field helix probably does not possess a preferred helicity because the DM-type antisymmetric interaction is absent for the present helimagnetic structure with $\mib{q} \parallel c$. 

The structure illustrated in Fig.~\ref{fig:Magst}(b) was constructed by assuming 
\begin{align}
\mib{\mu}(z) &= (m_a \cos qz \;\hat{\mib{x}} - m_b \sin qz \;\hat{\mib{y}}) \nonumber \\
 &+ A_{2q}(m_a \cos 2qz \;\hat{\mib{x}} - m_b \sin 2qz \;\hat{\mib{y}}) + m_{\text{F}} \hat{\mib{x}}  \;,
\end{align}
where $|m_a|=0.5 |m_b|$ was taken from Fig.~\ref{fig:MagFdep}(f), $A_{2q}=0.3$ from the $2q$ intensity, and a ferromagnetic component of $m_{\text{F}}=0.12$ was introduced. An equal-moment condition with $|\mib{\mu}|=1$ was then imposed. The resulting average ferromagnetic moment is 0.33, corresponding to 2.3 $\mu_{\text{B}}$/Eu, in agreement with the bulk magnetization measurement~\cite{Maurya16}. 

The equal-moment distorted helix shown in Fig.~\ref{fig:Magst}(b) may alternatively be described in the following way. 
Under the equal-moment assumption, the magnetic moments in the soliton-lattice state can be expressed as 
\begin{equation}
\mib{\mu}(z)=(\sin \phi(z), \cos \phi (z), 0) \;,
\end{equation}
where the phase $\phi(z)$ is written as 
\begin{equation}
\phi(z) = qz + \delta_1 \sin(qz) + \delta_2 \sin(2qz)\;.
\end{equation}
The zero-field helix is described by $\delta_1=\delta_2=0$. By introducing finite $\delta_1$ and $\delta_2$, the rotation angle become nonuniform, giving rise to the $2q$ Bragg peak.  
For $\delta_1=-0.08$ and $\delta_2=0.07$, the resulting average ferromagnetic moment is $0.31$, corresponding to a uniform magnetization of 2.2 $\mu_{\text{B}}$/Eu, in agreement with the experimental value. The Fourier transform of the magnetic structure yields $m_{qy}=1$ and $m_{qx}=0.60$, consistent with Fig.~\ref{fig:MagFdep}(f). This reflects the moment distribution shown in Fig.~\ref{fig:Magst}(c), which is no longer uniform but is preferentially oriented along the magnetic-field direction. 

The lock-in transition in phase III is reminiscent of the transition observed in the CSL state of the chiral helimagnet Yb(Ni$_{1-x}$Cu$_x$)$_3$Al$_9$ for $x=0.06$, where coupling to the lattice through Fermi-surface modification has been discussed~\cite{Okumura18}. 
Although EuRhGe$_3$ is not a chiral magnet, the application of a magnetic field induces $2q$, $3q$, and higher harmonic modulations. 
This feature is common to both EuRhGe$_3$ and the CSL state in Yb(Ni$_{1-x}$Cu$_x$)$_3$Al$_9$, suggesting a similar mechanism may be relevant. 
Since the $S=7/2$ spin state of Eu$^{2+}$ carries no orbital moment, crystal-field anisotropy is unlikely to be the primary origin of the coupling to the lattice. Furthermore, the fact that the magnetic anisotropy develops only below $T_{\text{N}}$ suggests that the conduction-electron contribution responsible for the RKKY interaction plays an important role both in determining the magnetic anisotropy and in driving the lock-in transition. 
Both effects may originate from the Fermi-surface structure reflecting the tetragonal symmetry of the crystal . 

\subsection{Phases I and IV}
Phase I corresponds to a conventional planar $xy$-fan phase, as inferred from Fig.~\ref{fig:pol8K}, and is realized as a high-field phase connected to the zero-field helical phase. 
The moments oscillate within the $ab$ plane and propagate along the $c$ axis. 
It should be noted, however, that an equal-moment fan structure necessarily contains a significant $2q$ Fourier component along the field direction ($a$ axis). Such a $2q$ peak has not been detected in phase I, nor in phase IV, indicating that the oscillating component along the $a$ axis is much smaller than expected for an equal-moment fan.  
Therefore, the actual magnetic structure in the fan phase may be described as 
\begin{align}
\mib{\mu}(z) &= -m_b \sin qz \;\hat{\mib{y}} - m_c \cos qz \;\hat{\mib{z}} \nonumber \\
 &+ \alpha \sqrt{1-(\mu_y^{\;2} + \mu_z^{\;2})} \;\hat{\mib{x}} \;,
\end{align}
where $m_c=0$ in the $xy$-fan structure of phase I. The coefficient $\alpha$ ($0 < \alpha < 1$) represents the modulation amplitude of the fan structure. The equal-moment circular fan corresponds to $\alpha=1$. 
From the experimental observation that the $2q$ intensity in the fan phase is less than $1/200$ of the primary $q$ intensity, the actual fan structure should resemble that illustrated in Fig.~\ref{fig:MagFdep}(f), where the oscillation of the $a$-axis component is suppressed and the magnitude of the ordered moment becomes modulated. 

With respect to phase IV, we conclude that a finite $c$-axis component appears in the fan structure, in which the dominant oscillation lies within the $ab$ plane but is accompanied by a weaker oscillation along the $c$ axis. This state has been predicted theoretically and is referred to as an elliptic-conical or $xyz$-fan structure, in contrast to the conventional planar $xy$-fan state in which the spins are confined to the $xy$ plane by magnetic anisotropy~\cite{Nagamiya62,Johnston19,Johnston17}. 
This elliptic-conical state arises from the tendency of magnetic moments to align perpendicular to the external field. 
For $H \parallel a$, in the absence of magnetic anisotropy, the moments would rotate in the $bc$-plane with a uniform $a$-axis component while propagating along the $c$ axis, forming a transverse conical structure. 
If the magnetic anisotropy is sufficiently strong, the helical order undergoes a direct transition to the planar $xy$-fan state without allowing the spins to flop out of the $ab$ plane. 
In EuRhGe$_3$, the magnetic anisotropy is strong enough to stabilize the zero-field helix within the $ab$ plane, yet sufficiently weak to allow the appearance of the intermediate elliptic-conical phase. As a result, the spins flop out of the $ab$ plane and  subsequently evolve into the conventional planar $xy$-fan state. 
To the best of our knowledge, EuRhGe$_3$ is the first helimagnet in which the full sequence from a planar helix to a planar $xy$-fan via an $xyz$-fan phase has been experimentally established. 

In Refs.~\citen{Nagamiya62} and \citen{Johnston19}, the phase diagrams were calculated at zero temperature as a function of magnetic anisotropy. 
In EuRhGe$_3$, the full sequence of phases is observed at low temperatures. At high temperatures, the planar helix undergoes a direct transition to the planar $xy$-fan without passing through the intermediate $xyz$-fan state. 
A possible explanation is that modulation of the ordered-moment magnitude becomes more favorable at elevated temperatures.  In such a situation, the Zeeman energy can be lowered by reducing the moment amplitude at unfavorable sites while keeping the spins within the $ab$ plane, thereby avoiding the anisotropy-energy cost associated with a spin flop.   
At lower temperatures, where the ordered moments are more fully developed, suppression of the moment amplitude becomes energetically less favorable. 
As a result, the Zeeman energy may overcome the anisotropy energy, allowing the moments to flop out of the $ab$ plane and thereby stabilizing the intermediate $xyz$-fan phase. 

Finally, we comment on an unresolved issue concerning the specific heat of EuRhGe$_3$. The $C(T)$ curve below $T_{\text{N}}$ exhibits two unusual features~\cite{Bednarchuk15,Maurya16,Kakihana17}. 
First, the jump at $T_{\text{N}}$ is approximately 14 J/(mol$\cdot$K), which is significantly smaller than the mean-field value of 21 J/(mol$\cdot$K) expected for an equal-moment magnetic structure. This reduction is not observed in other EuTGe$_3$ compounds with T=Pt, Pd, and Ni, which undergo a single magnetic transition. Second, the $C(T)$ curve is nearly flat at $T_{\text{N}}$ and does not exhibit the characteristic $\lambda$-type anomaly. 
Such a specific-heat profile resembles theoretical calculations for amplitude-modulated magnetic structures~\cite{Blanco91}. 
However, the present RXD results demonstrate that the zero-field helix in EuRhGe$_3$ is an equal-moment structure rather than an amplitude-modulated one, at least within the experimental accuracy of the present study. 
No indication of lattice distortion that would lower the tetragonal symmetry has been detected. 
Furthermore, the $T$-dependence of the order parameter is well described by the mean-field critical exponent $\beta=0.5$. 
At present, it is unclear how these observations can be reconciled. 
Understanding the origin of the anomalous specific-heat behavior remains an open issue for future investigation.

\section{Summary}
We have investigated the magnetic structure of the noncentrosymmetric helimagnet EuRhGe$_3$ by resonant X-ray diffraction in magnetic fields. 
The zero-field ordered state is an equal-amplitude planar helix in which the magnetic moments lie in the $ab$ plane and propagate along the $c$ axis with an incommensurate wave vector $\mib{q}=(0, 0, 0.809)$. 
When a magnetic field is applied along the $a$ axis, the planar-helix undergoes a direct transition to a planar $xy$-fan state at temperatures above 5 K. 
At lower temperatures below 5 K, however, two intermediate phases appear before the transition to the planar $xy$-fan state. 
First, the zero-field helix is distorted, accompanied by the development of a second harmonic $2q$ component, and finally undergoes a lock-in transition to a commensurate helix with $\mib{q}=(0, 0, 0.8)$. Subsequently, the oscillating component along the $a$ axis disappears and the spins flop out of the $ab$ plane, giving rise to an $xyz$-fan (elliptic conical) structure predicted theoretically. 
Thus, EuRhGe$_3$ exhibits a full sequence of field-induced phases from a planar helix to a lock-in helical phase, an spin flop $xyz$-fan phase, and finally a conventional planar $xy$-fan phase, before entering the field-induced ferromagnetic state. 

\begin{acknowledgment}
This work was supported by the JSPS Grant-in-Aid for Scientific Research (B) (No. JP20H01854) and JSPS Grant-in-Aid for Transformative Research Areas (Asymmetric Quantum Matters, No. JP23H04867). 
Synchrotron experiments were performed under the approval of the Photon Factory Program Advisory Committee (Nos. 2022G114 and 2024S2-002). 
\end{acknowledgment}

\appendix
\section{Scattering geometry}
\begin{figure}
\begin{center}
\includegraphics[width=8.5cm]{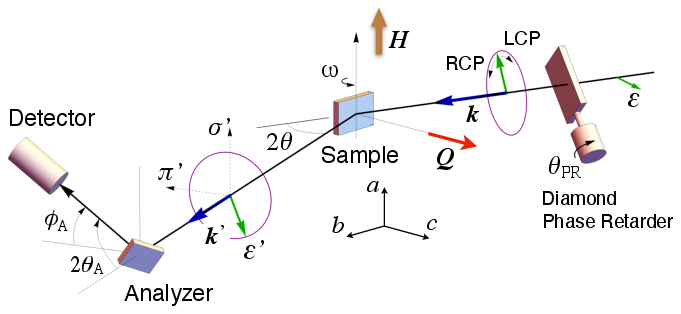}
\caption{(Color online) 
Scattering geometry of resonant X-ray diffraction in this work. 
$\mib{k}$ and $\mib{k}'$ denote the wave vectors of the incident and scattered X-rays, respectively. $2\theta$ and $2\theta_{\text{A}}$ are the scattering angles at the sample and analyzer crystal, respectively, and $\omega$ is the sample rocking angle. 
$\phi_{\text{A}}$ represents the detector angle measured from the horizontal plane. 
By inserting a phase-retarder system into the incident beam, the linear $\pi$ polarization from the synchrotron source can be converted into circular polarization by rotating the phase plate angle $\theta_{\text{PR}}$ about the diamond-(111) Bragg angle. 
In the present experiment, the phase retarder and polarization analyzer were not used simultaneously. 
When the phase retarder was employed, the detector was positioned before the analyzer. For polarization analysis, the phase retarder was removed from the beam path. 
}
\label{fig:ScattConfig}
\end{center}
\end{figure}

Figure \ref{fig:ScattConfig} shows the scattering geometry of the present RXD experiment.  
The incident X-rays from the synchrotron source are linearly polarized in the horizontal plane. 
To generate circularly polarized X-rays, a diamond-(111) crystal was used as a phase retarder. 
When the incident X-ray passes through the phase plate set near its Bragg angle $\theta_{\text{B}}$, 
a phase difference is introduced between the $\sigma$ and $\pi$ polarization components for the scattering plane tilted by $45^{\circ}$ with respect to the incident polarization. 
The phase difference is approximately given by $\alpha/(\theta_{\text{PR}} - \theta_{\text{B}})$, where $\alpha$ is an experimentally determined parameter of the phase plate.

\begin{figure}
\begin{center}
\includegraphics[width=7cm]{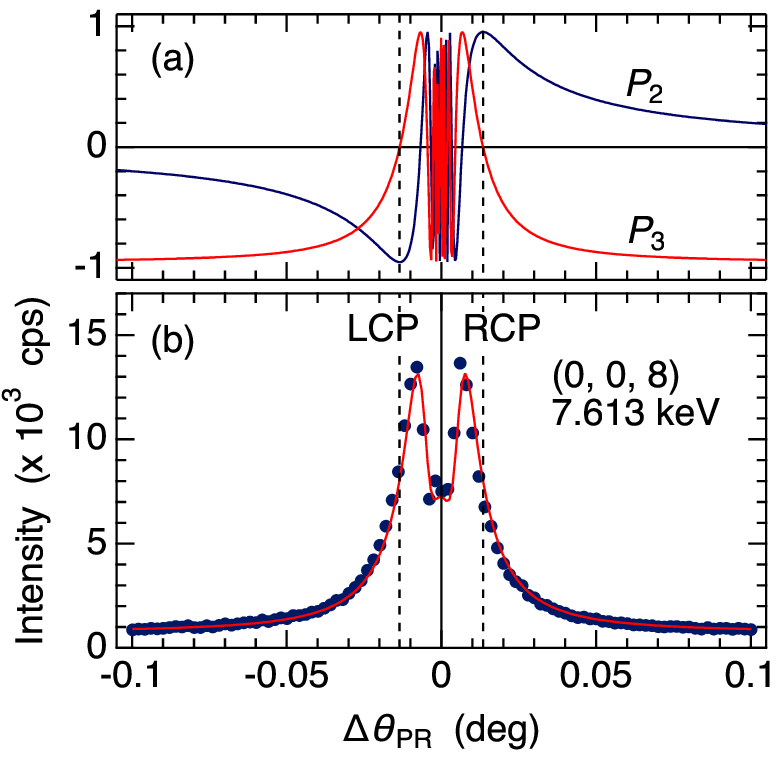}
\caption{(Color online) 
(a) $\Delta\theta_{\text{PR}}$ dependence of the Stokes parameters $P_2$ and $P_3$.  The parameters are given by 
$P_2=\sin (\alpha/\Delta\theta_{\text{PR}})$ and $P_3=-\cos (\alpha/\Delta\theta_{\text{PR}})$. 
(b) $\Delta\theta_{\text{PR}}$ dependence of the (8, 0, 0) fundamental Bragg-peak intensity. The solid line is a fit obtained by convolution with a Gaussian resolution function, from which $\alpha=0.0212$ was determined. The vertical dashed lines indicate the offset angles corresponding to the RCP and LCP states. 
}
\label{fig:P2P3param}
\end{center}
\end{figure}

The polarization state of the incident X-rays is described by the Stokes parameters $P_2$ and $P_3$,  
which represent the degrees of circular polarization ($+1$ for RCP and $-1$ for LCP) and linear polarization ($+1$ for $\sigma$ and $-1$ for $\pi$), respectively~\cite{Lovesey96}. 
In the geometry shown in Fig.~\ref{fig:ScattConfig}, where the scattering plane is horizontal,  the Stokes parameters are given by 
$P_2=\sin (\alpha/\Delta\theta_{\text{PR}})$ and $P_3=-\cos (\alpha/\Delta\theta_{\text{PR}})$, where $\Delta\theta_{\text{PR}} = \theta_{\text{PR}} - \theta_{\text{B}}$. 
By rotating the diamond phase plate about the Bragg angle $\theta_{\text{B}}$ of the (111) reflection, various mixtures of circular and linear polarization can be generated, as shown in Fig.~\ref{fig:P2P3param}(a). 

Figure \ref{fig:P2P3param}(b) shows the $\Delta\theta_{\text{PR}}$ dependence of the intensity of the fundamental Bragg reflection (0,0,8), which arises solely from Thomson scattering with $F_{\sigma\sigma'}=1$ and $F_{\pi\pi'}=\cos 2\theta$. 
The intensity is given by 
\begin{equation}
I \propto \Bigl( 1 - \frac{1-P_3}{2} \sin^2 2\theta \Bigr) \;. 
\end{equation}
The phase-plate parameter $\alpha=0.0212^{\circ}$ was determined by fitting the data with this expression. 
The resulting $\Delta\theta_{\text{PR}}$ dependences of $P_2$ and $P_3$ in the present set up are shown in Fig. \ref{fig:P2P3param}(a).  
In the fitting procedure, convolution with a Gaussian resolution function was included, leading to a suppression of the rapid oscillations near $\Delta\theta_{\text{PR}}\sim 0^{\circ}$

The general scattering intensity, including magnetic scattering, can be written as 
\begin{align}
I &= 
\frac{1}{2} \bigl(\, |F_{\sigma\sigma'}|^2 + |F_{\sigma\pi'}|^2 + |F_{\pi\sigma'}|^2 + |F_{\pi\pi'}|^2 \,\bigr) \nonumber \\
 &\;\;\;\; + P_1 \text{Re} \bigl\{\, F_{\pi\sigma'}^*F_{\sigma\sigma'} + F_{\pi\pi'}^*F_{\sigma\pi'} \,\bigr\} \nonumber \\
 &\;\;\;\; + P_2 \text{Im} \bigl\{\, F_{\pi\sigma'}^*F_{\sigma\sigma'} + F_{\pi\pi'}^*F_{\sigma\pi'} \,\bigr\} 
 \label{eq:CrossSec1} \\
 &\;\;\;\; +  \frac{1}{2} P_3\bigl(\, |F_{\sigma\sigma'}|^2 + |F_{\sigma\pi'}|^2 - |F_{\pi\sigma'}|^2 - |F_{\pi\pi'}|^2 \,\bigr) 
 \,. \nonumber
\end{align}
Therefore, the intensity for an incident beam characterized by the Stokes parameters $(P_1, P_2, P_3)$ can be written as
\begin{equation}
I = C_0 + C_1 P_1 + C_2 P_2 + C_3 P_3 \,,
\label{eq:CrossSec2}
\end{equation}
which provides a convenient fitting function for the $\Delta\theta_{\text{PR}}$ scans~\cite{Matsumura17}. 
In practice, $C_0$ serves as an overall scale factor, and $C_1$ does not contribute because $P_1=0$. 
Thus, only $C_2$ and $C_3$ remain as adjustable parameters. 
The helicity analysis shown in Fig.~\ref{fig:PRdep} was performed fitting the data with Eq.~(\ref{eq:CrossSec2}).


\bibliographystyle{jpsj}
\bibliography{EuRhGe3}

\end{document}